\documentclass[published]{JHEP3}
\JHEP{01(2002)006}

\usepackage{epsfig,graphicx}

\title{Rotating magnetic solution in three dimensional Einstein
gravity}

\author{\'Oscar J. C. Dias and Jos\'e P. S. Lemos\\
      	CENTRA, Departamento de F\'{\i}sica, Instituto Superior T\'ecnico\\
      	Av.\ Rovisco Pais 1, 1049-001 Lisboa, Portugal\\
	E-mail: \email{oscar@fisica.ist.utl.pt},
		\email{lemos@kelvin.ist.utl.pt}}

\abstract{We obtain the magnetic counterpart of the BTZ solution,
i.e., the rotating spacetime of a point source generating a magnetic
field in three dimensional Einstein gravity with a negative
cosmological constant. The static (non-rotating) magnetic solution was
found by Cl\'ement, by Hirschmann and Welch and by Cataldo and Salgado.  This paper
is an extension of their work in order to include (i) angular
momentum, (ii) the definition of conserved quantities (this is
possible since spacetime is asymptotically anti-de Sitter), (iii)
upper bounds for the conserved quantities themselves, and (iv) a new
interpretation for the magnetic field source.  We show that both the
static and rotating magnetic solutions have negative mass and that
there is an upper bound for the intensity of the magnetic field source
and for the value of the angular momentum.  The magnetic field source
can be interpreted not as a vortex but as being composed by a system
of two symmetric and superposed electric charges, one of the electric
charges is at rest and the other is spinning. The rotating magnetic
solution reduces to the rotating uncharged BTZ solution when the
magnetic field source vanishes.}

\accepted{January 8, 2002}
\received{December 5, 2001}

\begin{document}

\section{Introduction}

The three dimensional BTZ black hole solution of Ba\~nados, Teitelboim
and Zanelli~\cite{btz_PRL,btz_PRD} has been the subject of many
studies~\cite{Carlip}. The original article~\cite{btz_PRL} considers
the static and rotating uncharged BTZ black hole and the static
electrically charged BTZ black hole.  The extension to consider the
rotating electrically charged BTZ black hole (mass, angular momentum
and electric charge different from zero) has been done by
Cl\'ement~\cite{CL1} and by Mart\'{\i}nez, Teitelboim and
Zanelli~\cite{BTZ_Q}.  An extension to include a Brans-Dicke term has
been made by Dias and Lemos~\cite{OscarLemos}.

The static magnetic counterpart of the BTZ solution has also been
considered. Indeed Cl\'ement~\cite{CL1}, Hirschmann and Welch~\cite{HW} and Cataldo and
Salgado~\cite{Cat_Sal}, using different procedures, have found the
spacetime generated by a static point source of magnetic field in
three dimensional Einstein gravity with negative cosmological
constant, that reduces to the BTZ solution when the magnetic field
source vanishes.  This solution is horizonless and has a conical
singularity at the origin.

In this paper, we extend~\cite{HW,Cat_Sal} in order to include (i)
angular momentum, (ii) the definition of conserved quantities (mass,
angular momentum and electric charge), (iii) upper bounds for the
conserved quantities and (iv) a new interpretation for the magnetic
field source.  This rotating magnetic solution reduces to the rotating
uncharged BTZ solution when the magnetic field source vanishes.

The plan of this article is the following. In section~\ref{sec.2} we
study the static magnetic solution found in~\cite{HW,Cat_Sal} and its
properties.  Section~\ref{sec.3} is devoted to the rotating magnetic
solution.  The angular momentum is added in section~\ref{sec.3.1}
through a rotational Lorentz boost.  In section~\ref{sec.3.2} we
calculate the mass, angular momentum, and electric charge of both the
static and rotating solutions and in section~\ref{sec.3.3} we set
relations between the conserved charges.  The rotating magnetic
solution is written as a function of its hairs in
section~\ref{sec.3.4} and we show that it reduces to the rotating BTZ
solution when the magnetic source vanishes.  In section~\ref{sec.4} we
give a physical interpretation for the origin of the magnetic field
source.  Finally, in section~\ref{sec.5} we present the concluding
remarks.

\section{Static solution. Analysis of its general structure}
\label{sec.2}

\subsection{Static solution}
\label{sec.2.1}

Einstein gravity with a negative cosmological constant and a source of
magnetic field (or Einstein-Maxwell-anti de Sitter gravity) in three
dimensions can be characterized by the action
\begin{equation}
S=\frac{1}{4} \int d^3x \sqrt{-g} {\bigl [} R -2 \Lambda - F^{\mu
\nu}F_{\mu \nu} {\bigr ]}\,, 
\label{action}
\end{equation}
where $g$ is the determinant of the metric $g_{\mu\nu}$, $R$ the Ricci
scalar, $F_{\mu\nu}$ the electromagnetic tensor given in terms of the
vector potential $A_\mu$ by $F_{\mu\nu}=\partial_{\mu} A_\nu -
\partial_\nu A_\mu$, and $\Lambda$ is the cosmological constant.  In
this subsection we study the static point source solution of action
(\ref{action}) found by Cl\'ement~\cite{CL1}, by Hirschmann and Welch~\cite{HW} and by Cataldo
and Salgado~\cite{Cat_Sal}. The point source generates a gravitational
field and a magnetic field.  Written in the gauge presented
in~\cite{HW}, the static solution is given by the following metric and
vector potential 1-form
\begin{eqnarray} 
ds^2 &=& - (r^2/l^2 -m)dt^2 + [r^2+\chi^2_{\rm m} \ln{\mid r^2/l^2 - m
\mid}] d\varphi^2+
\nonumber \\&&
+ r^2(r^2/l^2 -m)^{-1} [r^2+\chi^2_{\rm m}\ln{\mid r^2/l^2-m
\mid}]^{-1}d r^2 \,,
\label{HW_0}  \\
A&=&\frac{1}{2} \chi_{\rm m} \ln{\mid r^2/l^2-m \mid} d \varphi \,,
\label{HW_Max}  
\end{eqnarray}
where $t$, $r$ and $\varphi$ are the time, the radial and the angular
coordinates, respectively, $\chi_{\rm m}$ is an integration constant
which measures the intensity of the magnetic field source and $l\equiv
-{1}/{\sqrt\Lambda}$ is the cosmological length.  This spacetime
reduces to the three dimensional BTZ black hole solution of Ba\~nados,
Teitelboim and Zanelli~\cite{btz_PRL,btz_PRD} when the magnetic source
vanishes.  The parameter $m$ is the mass of this uncharged solution.

The $g_{rr}$ function is negative for $r<r_+$ and positive for
$r>r_+$, where $r_+$ is such that
\begin{equation}
r_+^2 + \chi^2_{\rm m} \ln{\mid r_+^2/l^2-m \mid}=0 \,,  
\label{r+} 
\end{equation}
and the condition $r_+^2>ml^2$ is obeyed.  One might then be tempted
to say that the solution has an horizon at $r=r_+$ and consequently
that one is in the presence of a magnetically charged black
hole. However, this is not the case.  In fact, one first notices that
the metric components $g_{rr}$ and $g_{\varphi \varphi}$ are related
by $g_{\varphi \varphi}= [g_{rr}(r^2-ml^2)/(l^{2} r^{2})]^{-1}$.
Then, when $g_{rr}$ becomes negative (which occurs for $r<r_+$) so
does $g_{\varphi \varphi}$ and this leads to an apparent change of
signature from $+2$ to $-2$.  This strongly indicates~\cite{Hor_Hor}
that an incorrect extension is being used and that one should choose a
different continuation to describe the region $r<r_+$.  By introducing
a new radial coordinate, $\rho^2=r^2-r_+^2$, one obtains a spacetime
that is both null and timelike geodesically complete for $r\geq
r_+$~\cite{HW},
\begin{eqnarray}
ds^2 &=& -\frac{1}{l^2} (\rho^2+r_+^2 -m l^2)dt^2 + {\biggl
[}\rho^2+\chi^2_{\rm m} \ln{{\biggl (}1+ \frac{\rho^2}{r_+^2
-ml^2}{\biggr )}}{\biggr ]} d\varphi^2 +
\nonumber \\&&
+\, l^2 \rho^2(\rho^2+r_+^2-ml^2)^{-1} {\biggl [}\rho^2+\chi^2_{\rm
m}\ln{{\biggl (}1+ \frac{\rho^2}{r_+^2 -ml^2}{\biggr )}}{\biggr
]}^{-1} d \rho^2 \,,
\label{HW_1}  
\end{eqnarray}
where $0\leq \rho < \infty$. This static spacetime has no curvature
singularity, but it presents a conical geometry and, in particular, it
has a conical singularity at $\rho=0$ which can be removed if one
identifies $\varphi$ with the period $T_{\varphi}=2 \pi \nu$
where~\cite{HW} 
\begin{equation}
\nu=\frac{\exp{(\beta/2)}}{[1+\chi^2_{\rm m}\exp{(\beta)/l^2}]}\,,
\label{freq}
\end{equation}
and $\beta=r_+^2/\chi^2_{\rm m}$.  Near the origin, metric
(\ref{HW_1}) describes a spacetime which is locally flat but has a
conical singularity at $\rho=0$ with an angle deficit $\delta
\varphi=2\pi(1-\nu)$.

\subsection{Geodesic structure}
\label{sec.2.2}

We want to study the geodesic motion and, in particular, to confirm
that the spacetime described by (\ref{HW_1}) is both null and timelike
geodesically complete, i.e., that every null or timelike geodesic
starting from an arbitrary point either can be extended to infinite
values of the affine parameter along the geodesic or ends on a
singularity.  The equations governing the geodesics can be derived
from the lagrangian
\begin{equation}
{\cal{L}}=\frac{1}{2}g_{\mu\nu}\frac{dx^{\mu}}{d \tau}
\frac{dx^{\nu}}{d \tau}=-\frac{\delta}{2}\,,
\label{LAG)}  
\end{equation}
where $\tau$ is an affine parameter along the geodesic which, for a
timelike geodesic, can be identified with the proper time of the
particle along the geodesic. For a null geodesic one has $\delta=0$
and for a timelike geodesic $\delta=+1$. From the Euler-Lagrange
equations one gets that the generalized momenta associated with the
time coordinate and angular coordinate are constants: $p_t=E$ and
$p_{\varphi}=L$. The constant $E$ is related to the timelike Killing
vector $(\partial/\partial t)^{\mu}$ which reflects the time
translation invariance of the metric, while the constant $L$ is
associated to the spacelike Killing vector $(\partial/\partial
\varphi)^{\mu}$ which reflects the invariance of the metric under
rotation. Note that since the spacetime is not asymptotically flat,
the constants $E$ and $L$ cannot be interpreted as the energy and
angular momentum at infinity.

From the metric we can derive the radial geodesic, 
\begin{equation}
\dot{\rho}^2=-\frac{1}{g_{\rho\rho}} \frac{E^2 g_{\varphi \varphi}+L^2
g_{tt}} { g_{tt} g_{\varphi \varphi} } -\frac{\delta}{g_{\rho\rho}}
\,.
\label{GEOD_1}
\end{equation}
Using the two useful relations $g_{tt} g_{\varphi
\varphi}=-\rho^2/g_{\rho\rho}$ and $g_{\varphi \varphi}=[g_{\rho\rho}
(\rho^2+r_+^2-ml^2) /(l^2 \rho^2)]^{-1}$, we can write
eq.~(\ref{GEOD_1}) as
\begin{equation}
\rho^2 \dot{\rho}^2= {\biggl [} \frac{l^2 E^2}{\rho^2 + r_+^2-ml^2}
-\delta {\biggr ]} \frac{\rho^2}{g_{\rho \rho}} +L^2 g_{tt} \,.
\label{Geod_1}
\end{equation}
(i) Null geodesics ($\delta=0$) $-$ Noticing that $1/g_{\rho\rho}$ is
always positive for $\rho>0$ and zero for $\rho=0$, and that $g_{tt}$
is always negative we conclude the following about the null geodesic
motion.  The first term in (\ref{Geod_1}) is positive (except at
$\rho=0$ where it vanishes), while the second term is always
negative. We can then conclude that spiraling ($L \neq 0$) null
particles coming in from an arbitrary point are scattered at the
turning point $\rho_{\rm tp} > 0$ and spiral back to infinity. If the
angular momentum L of the null particle is zero it hits the origin
(where there is a conical singularity) with vanishing velocity.  (ii)
Timelike geodesics ($\delta=+1$) $-$ Timelike geodesic motion is
possible only if the energy of the particle satisfies $E >
(r_+^2-ml^2)^{1/2}/l$. In this case, spiraling timelike particles are
bounded between two turning points that satisfy $\rho_{\rm tp}^{\rm a}
> 0$ and $\rho_{\rm tp}^{\rm b} < \sqrt{l^2(E^2+m) - r_+^2}$, with
$\rho_{\rm tp}^{\rm b} \geq \rho_{\rm tp}^{\rm a}$.  When the timelike
particle has no angular momentum ($L=0$) there is a turning point
located at $\rho_{\rm tp}^{\rm b}=\sqrt{l^2(E^2+m) - r_+^2}$ and it
hits the conical singularity at the origin $\rho=0$.  Hence, we
confirm that the spacetime described by eq. (\ref{HW_1}) is both
timelike and null geodesically complete.

\section{Rotating magnetic solution}
\label{sec.3}

\subsection{Addition of angular momentum}
\label{sec.3.1}

Now, we want to endow the spacetime solution (\ref{HW_1}) with a
global rotation, i.e., we want to add angular momentum to the
spacetime.  In order to do so we perform the following rotation boost
in the $t$-$\varphi$ plane (see e.g.~\cite{BTZ_Q,OscarLemos,HorWel})
\begin{eqnarray}
t &\mapsto& \gamma t-l \omega \varphi \,,
\nonumber  \\
\varphi &\mapsto& \gamma \varphi-\frac{\omega}{l} t \,,
\label{TRANSF_J_HW} 
\end{eqnarray}
where $\gamma$ and $\omega$ are constant parameters.  Substituting
(\ref{TRANSF_J_HW}) into (\ref{HW_1}) and (\ref{HW_Max}) we obtain the
stationary spacetime generated by a magnetic source
\begin{eqnarray} 
ds^2 &=& -\frac{1}{l^2} {\biggl [}(\gamma^2-\omega^2)\rho^2
+\gamma^2(r_+^2 -m l^2) -\omega^2 \chi^2_{\rm m} \ln{{\biggl (}1+
\frac{\rho^2}{r_+^2 -ml^2}{\biggr )}}{\biggr ]}dt^2-
\nonumber \\&&
-\frac{\gamma \omega}{l} {\biggl [} -(r_+^2 -m l^2) + \chi^2_{\rm m}
\ln{{\biggl (}1+ \frac{\rho^2}{r_+^2 -ml^2}{\biggr )}}{\biggr ]} 2 dt
d \varphi+
\nonumber \\&&
+ l^2 \rho^2(\rho^2+r_+^2-ml^2)^{-1} {\biggl [}\rho^2+\chi^2_{\rm
m}\ln{{\biggl (}1+ \frac{\rho^2}{r_+^2 -ml^2}{\biggr )}}{\biggr
]}^{-1} d \rho^2 +
\nonumber \\&&
+ {\biggl [}(\gamma^2-\omega^2)\rho^2 -\omega^2(r_+^2 -m l^2)+\gamma^2
\chi^2_{\rm m} \ln{{\biggl (}1+ \frac{\rho^2}{r_+^2 -ml^2}{\biggr
)}}{\biggr ]} d\varphi^2 \,,
\label{HW_Rot}  \\
A &=& -\frac{\omega}{l}A(\rho)dt + \gamma A(\rho) d \varphi \,,   
\label{HW_Max_Rot}  
\end{eqnarray}
with $A(\rho)= \chi_{\rm m} \ln{[(\rho^2+r_+^2)/l^2-m]}/2$.  We set
$\gamma^2-\omega^2=1$ because in this way when the angular momentum
vanishes ($\omega=0$) we have $\gamma=1$ and so we recover the static
solution.

Solution (\ref{HW_Rot}) represents a magnetically charged stationary
spacetime and also solves the three dimensional Einstein-Maxwell-anti
de Sitter gravity action (\ref{action}).  Transformations
(\ref{TRANSF_J_HW}) generate a new metric because they are not
permitted global coordinate transformations
\cite{Stachel}. Transformations (\ref{TRANSF_J_HW}) can be done
locally, but not globally. Therefore, the metrics (\ref{HW_1}) and
(\ref{HW_Rot}) can be locally mapped into each other but not globally,
and as such they are distinct.

Cl\'ement~\cite{CL1}, using a procedure, and Chen~\cite{Chen}, through
the application of $T$-duality to~\cite{HW} have written a rotating
metric. However, the properties of the spacetime were not studied.

\subsection{Mass, angular momentum and electric charge
of the solutions}\label{sec.3.2}

Both the static and rotating solutions are asymptotically anti-de
Sitter.  This fact allows us to calculate the mass, angular momentum
and the electric charge of the static and rotating solutions.  To
obtain these quantities we apply the formalism of Regge and
Teitelboim~\cite{Regge} (see also~\cite{BTZ_Q,OscarLemos}).  We first
write (\ref{HW_Rot}) in the canonical form involving the lapse
function $N^0(\rho)$ and the shift function $N^{\varphi}(\rho)$
\begin{equation}
ds^2 = - (N^0)^2 dt^2 + \frac{d\rho^2}{f^2} +
H^2(d\varphi+N^{\varphi}dt)^2 \,, 
\label{MET_CANON}
\end{equation}
where $f^{-2}=g_{\rho\rho}$, $H^2=g_{\varphi \varphi}$, $H^2
N^{\varphi}=g_{t \varphi}$ and $(N^0)^2-H^2(N^{\varphi})^2=g_{tt}$.
Then, the action can be written in the hamiltonian form as a function
of the energy constraint ${\cal{H}}$, momentum constraint
${\cal{H}}_{\varphi}$ and Gauss constraint $G$
\begin{eqnarray}
S &=& -\int dt d^2x[N^0 {\cal{H}}+N^{\varphi} {\cal{H}_{\varphi}} +
A_{t} G]+ {\cal{B}}
\nonumber \\
&=& -\Delta t \int d\rho N \nu {\biggl [} \frac{2 \pi^2}{H^3}
+2f(fH_{,\rho})_{,\rho} +\frac{H}{l^2} +\frac{2H}{f}(E^2+B^2){\biggr
]}+
\nonumber \\&&
+ \Delta t \int d\rho N^{\varphi}\nu{\biggl [} {\bigl (}2 \pi {\bigr
)}_{,\rho} +\frac{4H}{f}E^{\rho}B{\biggr ]} + \Delta t \int d \rho A_t
\nu {\biggl [}-\frac{4H}{f} \partial_{\rho} E^\rho{\biggr ]}
+{\cal{B}} \,,
\label{ACCAO_CANON}
\end{eqnarray}
where $N=\frac{N^0}{f}$, $\pi \equiv {\pi_{\varphi}}^{\rho}=
-\frac{fH^3 (N^{\varphi})_{,\rho}}{2N^0}$ (with $\pi^{\rho \varphi}$
being the momentum conjugate to $g_{\rho \varphi}$), $E^{\rho}$ and
$B$ are the electric and magnetic fields and ${\cal{B}}$ is a boundary
term.  The factor $\nu$ comes from the fact that, due to the angle
deficit, the integration over $\varphi$ is between $0$ and $2 \pi\nu$.
Upon varying the action with respect to $f(\rho)$, $H(\rho)$,
$\pi(\rho)$ and $E^{\rho}(\rho)$ one picks up additional surface
terms.  Indeed,
\begin{eqnarray}
\delta S &=& - \Delta t N \nu{\biggl [}H_{,\rho} \delta f^2
-(f^2)_{,\rho}\delta H +2f^2 \delta (H_{,\rho}) {\biggr ]}+
\nonumber \\&&
+\Delta t N^{\varphi} [2 \nu \delta \pi] + \Delta t A_t {\biggl [} -
\nu\frac{4H}{f} \delta E^{\rho}{\biggr ]} + \delta {\cal{B}}+
\nonumber \\&&
+(\mbox{terms vanishing when the equations of motion hold})\,.
\label{DELTA_ACCAO}
\end{eqnarray}
In order that the Hamilton's equations are satisfied, the boundary
term ${\cal{B}}$ has to be adjusted so that it cancels the above
additional surface terms.  More specifically one has
\begin{equation}
\delta {\cal{B}} = -\Delta t N \delta M +\Delta t N^{\varphi}\delta J
+ \Delta t A_t \delta Q_{\rm e} \,,
\label{DELTA_B}
\end{equation}
where one identifies $M$ as the mass, $J$ as the angular momentum and
$Q_{\rm e}$ as the electric charge since they are the terms conjugate
to the asymptotic values of $N$, $N^{\varphi}$ and $A_t$,
respectively.

\pagebreak[3]

To determine the mass, the angular momentum and the electric charge of
the solutions one must take the spacetime that we have obtained and
subtract the background reference spacetime contribution, i.e., we
choose the energy zero point in such a way that the mass, angular
momentum and charge vanish when the matter is not present.

Now, note that~(\ref{HW_Rot}) has an asymptotic metric given by
\begin{equation}
-\frac{\gamma^2-\omega^2}{l^2}\rho^2 dt^2+ \frac{l^2}{\rho^2}d \rho^2+
(\gamma^2-\omega^2) \rho^2 d \varphi^2 \,, 
\label{ANTI_SITTER}
\end{equation}
where $\gamma^2-\omega^2 =1$ so, it is asymptotically an anti-de
Sitter spacetime.  The anti-de Sitter spacetime is also the background
reference spacetime, since the metric (\ref{HW_Rot}) reduces
to~(\ref{ANTI_SITTER}) if the matter is not present ($m=0$ and
$\chi_{\rm m}=0$).

Taking the subtraction of the background reference spacetime into
account we have that the mass, angular momentum and electric charge
are given by
\begin{eqnarray}
M &=& \nu {\bigl [}-H_{,\rho}(f^2-f^2_{\rm ref})
+(f^2)_{,\rho}(H-H_{\rm ref}) -2f^2 (H_{,\rho}-H_{,\rho}^{\rm ref})
{\bigr ]}\,,
\nonumber \\
J &=& -2\nu (\pi-\pi_{\rm ref}) \,,
\nonumber \\
Q_{\rm e} &=& \frac{4H}{f} \nu (E^{\rho}-E^{\rho}_{\rm ref}) \,.
\label{MQ_GERAL}
\end{eqnarray}  
After taking the asymptotic limit, $\rho \rightarrow +\infty$, we
finally have that the mass and angular momentum are
\begin{eqnarray}
M &=& \nu {\bigl [}(\gamma^2+\omega^2)(m - r_+^2/l^2) -2 \chi^2_{\rm
m}/l^2 {\bigr ]} + {\rm Div_M}(\chi_{\rm m},\rho) \,,
\label{M} \\
J &=& 2 \nu \gamma \omega(ml^2-r_+^2-\chi^2_{\rm m})/l + {\rm
Div_J}(\chi_{\rm m},\rho)\,,
\label{J}
\end{eqnarray}  
where ${\rm Div_M}(\chi_{\rm m},\rho)$ and ${\rm Div_J}(\chi_{\rm
m},\rho)$ are logarithmic terms proportional to the magnetic source
$\chi_{\rm m}$ that diverge as $\rho \rightarrow +\infty$ (see
also~\cite{Chan}).  The presence of these kind of divergences in the
mass and angular momentum is a usual feature present in charged
solutions. They can be found for example in the electrically charged
point source solution~\cite{Deser_Maz}, in the electrically charged
BTZ black hole~\cite{BTZ_Q} and in the electrically charged black
holes of three dimensional Brans-Dicke
gravity~\cite{OscarLemos}. Following~\cite{BTZ_Q,OscarLemos,Deser_Maz}
the divergences on the mass can be treated as follows. One considers a
boundary of large radius $\rho_0$ involving the system. Then, one sums
and subtracts ${\rm Div_M}(\chi_{\rm m},\rho_0)$ to (\ref{M}) so that
the mass~(\ref{M}) is now written~as
\begin{equation}
M = M(\rho_0)+ [{\rm Div_M}(\chi_{\rm m},\rho)- {\rm Div_M}(\chi_{\rm
m},\rho_0)] \,, 
\label{M0_0}
\end{equation}  
where $M(\rho_0)=M_0+{\rm Div_M}(\chi_{\rm m},\rho_0)$, i.e.,
\begin{equation}
M_0=M(\rho_0)-{\rm Div_M}(\chi_{\rm m},\rho_0)\,.  
\label{M0_0_v2}
\end{equation}  
The term between brackets in (\ref{M0_0}) vanishes when $\rho
\rightarrow \rho_0$. Then $M(\rho_0)$ is the energy within the radius
$\rho_0$. The difference between $M(\rho_0)$ and $M_0$ is $-{\rm
Div_M}(\chi_{\rm m},\rho_0)$ which is interpreted as the
electromagnetic energy outside $\rho_0$ apart from an infinite
constant which is absorbed in $M(\rho_0)$. The sum (\ref{M0_0_v2}) is
then independent of $\rho_0$, finite and equal to the total mass.  In
practice the treatment of the mass divergence amounts to forgetting
about $\rho_0$ and take as zero the asymptotic limit: $\lim {\rm
Div_M}(\chi_{\rm m},\rho)=0$.

To handle the angular momentum divergence, one first notices that the
asymptotic limit of the angular momentum per unit mass $(J/M)$ is
either zero or one, so the angular momentum diverges at a rate slower
or equal to the rate of the mass divergence.  The divergence on the
angular momentum can then be treated in a similar way as the mass
divergence. So, one can again consider a boundary of large radius
$\rho_0$ involving the system. Following the procedure applied for the
mass divergence one concludes that the divergent term $-{\rm
Div_J}(\chi_{\rm m},\rho_0)$ can be interpreted as the electromagnetic
angular momentum outside $\rho_0$ up to an infinite constant that is
absorbed in $J(\rho_0)$.

Hence, in practice the treatment of both the mass and angular
divergences amounts to forgetting about $\rho_0$ and take as zero the
asymptotic limits: $\lim {\rm Div_M}(\chi_{\rm m},\rho)=0$ and $\lim
{\rm Div_J}(\chi_{\rm m},\rho)=0$ in (\ref{M}) and (\ref{J}).

Now, we calculate the electric charge of the solutions.  To determine
the electric field we must consider the projections of the Maxwell
field on spatial hypersurfaces.  The normal to such hypersurfaces is
$n^{\nu}=(1/N^0,0,-N^{\varphi}/N^0)$ and the electric field is given
by $E^{\mu}=g^{\mu \sigma}F_{\sigma \nu}n^{\nu}$.  Then, from
(\ref{MQ_GERAL}), the electric charge is
\begin{equation}
Q_{\rm e}=-\frac{4Hf}{N^0} \nu (\partial_{\rho}A_t-N^{\varphi}
\partial_{\rho} A_{\varphi})= 2 \nu \frac{\omega}{l}\chi_{\rm m} \,.
\label{CARGA}
\end{equation}
Note that the electric charge is proportional to $\omega \chi_{\rm
m}$.  Since in three dimensions the magnetic field is a scalar (rather
than a vector) one cannot use Gauss's law to define a conserved
magnetic charge.  In the next section we will propose a physical
interpretation for the origin of the magnetic field source and discuss
the result obtained in (\ref{CARGA}).

The mass, angular momentum and electric charge of the static solutions
can be obtained by putting $\gamma=1$ and $\omega=0$ on the above
expressions [see (\ref{TRANSF_J_HW})].

\subsection{Relations between the conserved charges}
\label{sec.3.3}

Now, we want to cast the metric (\ref{HW_Rot}) in terms of $M$, $J$,
$Q_{\rm e}$ and $\chi_{\rm m}$.  We can use (\ref{M}) and (\ref{J}) to
solve a quadratic equation for $\gamma^2$ and $\omega^2$. It gives two
distinct sets of solutions
\begin{eqnarray}
\gamma^2&=&\frac{Ml^2 + 2\chi_{\rm m}^2}{2(ml^2-r_+^2)}
\frac{(2-\Omega)}{\nu} \,, \qquad \omega^2= \frac{Ml^2 + 2\chi_{\rm
m}^2}{2(ml^2-r_+^2)} \frac{\Omega}{\nu} \,,
\label{DUAS_HW}\\
\gamma^2&=&\frac{Ml^2 + 2\chi_{\rm m}^2}{2(ml^2-r_+^2)}
\frac{\Omega}{\nu} \,, \qquad \omega^2= \frac{Ml^2 + 2\chi_{\rm
m}^2}{2(ml^2-r_+^2)} \frac{(2-\Omega)}{\nu} \,,
\label{DUAS_ERR_HW}
\end{eqnarray}
where we have defined a rotating parameter $\Omega$, which ranges
between $0 \leq \Omega < 1$, as
\begin{equation}
\Omega \equiv 1- \sqrt{1-\frac{(ml^2-r_+^2)^2} {(Ml^2 + 2\chi_{\rm
m}^2)^2} \frac{l^2 J^2}{(ml^2-r_+^2-\chi^2_{\rm m})^2}} \,.
\label{OMEGA_HW}
\end{equation}
When we take $J=0$ (which implies $\Omega=0$), (\ref{DUAS_HW}) gives
$\gamma \neq 0$ and $\omega= 0$ while (\ref{DUAS_ERR_HW}) gives the
nonphysical solution $\gamma=0$ and $\omega \neq 0$ which does not
reduce to the static original metric.  Therefore we will study the
solutions found from (\ref{DUAS_HW}).  The condition that $\Omega$
remains real imposes a restriction on the allowed values of the
angular momentum
\begin{equation}
l^2 J^2 \leq \frac{(ml^2-r_+^2-\chi^2_{\rm m})^2}{(ml^2-r_+^2)^2}
(Ml^2 + 2\chi_{\rm m}^2)^2\,.  
\label{Rest_OMEGA_HW}
\end{equation}

The condition $\gamma^2-\omega^2=1$ allows us to write $r_+^2-ml^2$ as
a function of $M$, $\Omega$ and $\chi_{\rm m}$,
\begin{equation}
r_+^2-ml^2= (Ml^2 + 2\chi_{\rm m}^2) (\Omega-1)/\nu \,.
\label{b_HW} 
\end{equation}
This relation allows us to achieve interesting conclusions about the
values that the parameters $M$, $\chi_{\rm m}$ and $J$ can
have. Indeed, if we replace (\ref{b_HW}) into (\ref{DUAS_HW}) we get
\begin{equation}
\gamma^2=\frac{(2-\Omega)}{2(1-\Omega)}\,, \qquad \omega^2=
\frac{\Omega}{2(1-\Omega)} \,.
\label{DUAS_HW_v2}
\end{equation}
Since $\Omega$ ranges between $0 \leq \Omega < 1$, we have
$\gamma^2>0$ and $\omega^2>0$.  Besides, one has that $r_+^2>ml^2$ and
$\nu>0$ so from (\ref{M0_0}) we conclude that both the static and
rotating solutions have negative mass. Therefore, from now one,
whenever we refer to the mass of the solution we will set
\begin{equation}
M=-|M|\,,
\label{mod_m}
\end{equation}
unless otherwise stated.

Looking again to (\ref{b_HW}) we can also conclude that one must have
\begin{equation}
\chi_{\rm m}^2<\frac{|M|l^2}{2}\,,
\label{v_max_q}
\end{equation}
i.e., there is an upper bound for the intensity of the magnetic field
strength.

From (\ref{J}) we also see that the angular momentum is always
negative indicating that the angular momentum and the angular
velocity, $\omega$, have opposite directions. This is the expected
result since $J$ is the inertial momentum times the angular velocity
and the inertial momentum is proportional to the mass which is
negative.  Introducing (\ref{b_HW}) into (\ref{Rest_OMEGA_HW}) we find
an upper bound for the angular momentum
\begin{equation}
|J|\leq |M|l^2-2 \chi_{\rm m}^2+\nu\chi_{\rm m}^2/(1-\Omega)\,.
\label{J_max}
\end{equation}
Note that from (\ref{OMEGA_HW}) we can get the precise value of $J$ as
a function of $M$, $\Omega$ and $\chi_{\rm m}$.

Finally, we remark that the auxiliary equations (\ref{r+}),
(\ref{freq}) and (\ref{DUAS_HW_v2}) allow us to define the auxiliary
parameters $r_+$, $\nu$ and $m$ as a function of the hairs $M$,
$\Omega$ and $\chi_{\rm m}$.

\subsection{The rotating magnetic solution}
\label{sec.3.4}

We are now in position to write the stationary spacetime
(\ref{HW_Rot}) generated by a source of magnetic field in three
dimensional Einstein-Maxwell-anti de Sitter gravity as a function of
its hairs,
\begin{eqnarray} 
ds^2 &=& -\frac{1}{l^2} {\biggl [}\rho^2 +\frac{1}{2\nu}(|M|l^2-
2\chi_{\rm m}^2)(2-\Omega) -\frac{Q_{\rm e}^2}{4 \nu} \ln{{\biggl (}1+
\frac{\nu \rho^2}{(|M|l^2- 2\chi_{\rm m}^2)(1-\Omega)} {\biggr
)}}{\biggr ]}dt^2+\quad
\nonumber \\&&
+\frac{J}{\nu}\frac{ (|M|l^2 - 2\chi_{\rm m}^2)(\Omega-1) + \nu
\chi^2_{\rm m} \ln{{\biggl (}1+ \frac{\nu\rho^2}{(|M|l^2 - 2\chi_{\rm
m}^2)(1-\Omega)} {\biggr )}}}{(|M|l^2 - 2\chi_{\rm m}^2)(1-\Omega)
+\nu \chi_{\rm m}^2} dt d \varphi+
\nonumber \\&&
+ \frac{l^2 \rho^2 {\biggl [}\rho^2+\chi^2_{\rm m}\ln{{\biggl (}1+
\frac{\nu \rho^2}{(|M|l^2 - 2\chi_{\rm m}^2)(1-\Omega)} {\biggr
)}}{\biggr ]}^{-1}}{\rho^2+(|M|l^2 - 2\chi_{\rm m}^2) (1-\Omega)/\nu}
d \rho^2+
\nonumber \\&&
+ {\biggl [}\rho^2 -(|M|l^2 - 2\chi_{\rm m}^2)\frac{\Omega}{2 \nu}+
\frac{2-\Omega}{1-\Omega} \frac{\chi^2_{\rm m}}{2} \ln{{\biggl (}1+
\frac{\nu \rho^2}{(|M|l^2 - 2\chi_{\rm m}^2)(1-\Omega)} {\biggr
)}}{\biggr ]} d\varphi^2 \,,\quad
\label{HW_Rot_Fim}  
\end{eqnarray} 
as well as the vector potential 1-form (\ref{HW_Max_Rot})
\begin{equation}
A=\frac{2}{\sqrt{1-\Omega}}{\biggl [}-\frac{\sqrt{\Omega}}{l}A(\rho)dt
+\sqrt{2-\Omega} A(\rho) d \varphi{\biggr ]} \,,
\label{HW_Max_Rot_Fim}  
\end{equation}
with $A(\rho)=(\chi_{\rm m}/2) \ln{[\rho^2/l^2+(|M|-2\chi_{\rm
m}^2/l^2) (1-\Omega)/\nu]}$.

If we set $\Omega=0$ (and thus $J=0$ and $Q_{\rm e}=0$) we recover the
static solution (\ref{HW_1}) [see (\ref{TRANSF_J_HW})]. Finally if we
set $\chi_{\rm m}=0$ (and so $\nu=1$) one gets
\begin{equation}
ds^2 = -\frac{1}{l^2} {\biggl [}\rho^2 -Ml^2\frac{2-\Omega}{2} {\biggr
]}dt^2 -J dt d \varphi + \frac{l^2} {\rho^2-Ml^2 (1-\Omega)} d \rho^2
+ {\biggl [}\rho^2 +Ml^2 \frac{\Omega}{2}{\biggr ]} d\varphi^2 \,,
\label{BTZ_0}  
\end{equation}
where we have dropped the absolute value of $M$ since now the mass can
be positive.  This is the rotating uncharged BTZ solution written,
however, in an unusual gauge.  To write it in the usual gauge we apply
to (\ref{BTZ_0}) the radial coordinate transformation
\begin{equation}
\rho^2=R^2- Ml^2 \frac{\Omega}{2} \:\:\:\Rightarrow \:\:\:
d\rho^2=\frac{R^2}{R^2-Ml^2 \Omega /2} dR^2
\label{transf_BTZ}  
\end{equation}
and use the relation $J^2=\Omega(2-\Omega)Ml^2$ [see (\ref{OMEGA_HW})]
to obtain
\begin{equation}
ds^2 = - {\biggl (}\frac{R^2}{l^2} -M{\biggr )}dt^2 -J dt d \varphi +
{\biggl (}\frac{R^2}{l^2}-M+\frac{J^2}{4R^2} {\biggr )}^{-1} d R^2 +
R^2 d\varphi^2 \,.
\label{BTZ}  
\end{equation}
So, as expected, (\ref{HW_Rot_Fim}) reduces to the rotating uncharged
BTZ solution~\cite{btz_PRL,btz_PRD} when the magnetic field source
vanishes.

\subsection{Geodesic structure}
\label{sec.3.5}

The geodesic structure of the rotating spacetime is similar to the
static spacetime (see section II.2), although there are now direct
(corotating with $L>0$) and retrograde (counter-rotating with $L<0$)
orbits.  The most important result that spacetime is geodesically
complete still holds for the stationary spacetime.

\section{Physical interpretation of the magnetic source}
\label{sec.4}

When we look back to the electric charge given in~(\ref{CARGA}), we
see that it is zero when $\omega=0$, i.e., when the angular momentum
$J$ of the spacetime vanishes.  This is expected since in the static
solution we have imposed that the electric field is zero ($F_{12}$ is
the only non-null component of the Maxwell tensor).

Still missing is a physical interpretation for the origin of the
magnetic field source.  The magnetic field source is not a
Nielson-Oleson vortex solution since we are working with the Maxwell
theory and not with an abelian-Higgs model.  We might then think that
the magnetic field is produced by a Dirac point-like
monopole. However, this is not also the case since a Dirac monopole
with strength $g_{\rm m}$ appears when one breaks the Bianchi
identity~\cite{MeloNeto}, yielding $\partial_{\mu} (\sqrt{-g}
\tilde{F}^{\mu})= g_{\rm m} \delta^2 (\vec{x})$ (where
$\tilde{F}^{\mu}=\epsilon^{\mu \nu \gamma}F_{\nu \gamma}/2$ is the
dual of the Maxwell field strength), whereas in this work we have that
$\partial_{\mu} (\sqrt{-g} \tilde{F}^{\mu})=0$.  Indeed, we are
clearly dealing with the Maxwell theory which satisfies Maxwell
equations and the Bianchi identity
\begin{eqnarray}
\frac{1}{\sqrt{-g}}\partial_{\nu}(\sqrt{-g}F^{\mu \nu})
&=&\frac{\pi}{2}\frac{1}{\sqrt{-g}} j^{\mu} \,,
\label{Max_j} \\
\partial_{\mu} (\sqrt{-g} \tilde{F}^{\mu})&=&0 \,,
\label{Max_bianchi}
\end{eqnarray}  
respectively. In (\ref{Max_j}) we have made use of the fact that the
general relativistic current density is $1/\sqrt{-g}$ times the
special relativistic current density $j^{\mu}=\sum q
\delta^2(\vec{x}-\vec{x}_0)\dot{x}^{\mu}$.

We then propose that the magnetic field source can be interpreted as
composed by a system of two symmetric and superposed electric charges
(each with strength $q$). One of the electric charges is at rest with
positive charge (say), and the other is spinning with an angular
velocity $\dot{\varphi}_0$ and negative electric charge.  Clearly,
this system produces no electric field since the total electric charge
is zero and the magnetic field is produced by the angular electric
current.  To confirm our interpretation, we go back to
eq. (\ref{Max_j}).  In our solution, the only non-vanishing component
of the Maxwell field is $F^{\varphi \rho}$ which implies that only
$j^{\varphi}$ is not zero. According to our interpretation one has
$j^{\varphi}=q \delta^2(\vec{x}-\vec{x}_0)\dot{\varphi}$, which one
inserts in eq. (\ref{Max_j}).  Finally, integrating over $\rho$ and
$\varphi$ we have
\begin{equation}
\chi_{\rm m} \propto q \dot{\varphi}_0 \,.
\label{Q_mag}
\end{equation} 
So, the magnetic source strength, $\chi_{\rm m}$, can be interpreted
as an electric charge $q$ times its spinning velocity.

Looking again to the electric charge given in (\ref{CARGA}), one sees
that after applying the rotation boost in the $t$-$\varphi$ plane to
endow the initial static spacetime with angular momentum, there
appears a net electric charge.  This result was already expected since
now, besides the scalar magnetic field ($F_{\rho \varphi} \neq 0$),
there is also an electric field ($F_{t \rho} \neq 0$) [see
(\ref{HW_Max_Rot_Fim})].  A physical interpretation for the appearance
of the net electric charge is now needed. To do so, we return to the
static spacetime.  In this static spacetime there is a static positive
charge and a spinning negative charge of equal strength at the center.
The net charge is then zero. Therefore, an observer at rest ($S$) sees
a density of positive charges at rest which is equal to the density of
negative charges that are spinning. Now, we perform a local rotational
boost $t'= \gamma t-l \omega \varphi$ and $\varphi' = \gamma
\varphi-\frac{\omega}{l} t\:$ to an observer ($S'$) in the static
spacetime, so that $S'$ is moving relatively to $S$.  This means that
$S'$ sees a different charge density since a density is a charge over
an area and this area suffers a Lorentz contraction in the direction
of the boost.  Hence, the two sets of charge distributions that had
symmetric charge densities in the frame $S$ will not have charge
densities with equal magnitude in the frame $S'$. Consequently, the
charge densities will not cancel each other in the frame $S'$ and a
net electric charge appears.  This was done locally. When we turn into
the global rotational Lorentz boost of eqs. (\ref{TRANSF_J_HW}) this
interpretation still holds.  The local analysis above is similar to
the one that occurs when one has a copper wire with an electric
current and we apply a translation Lorentz boost to the wire: first,
there is only a magnetic field but, after the Lorentz boost, one also
has an electric field.  The difference is that in the present
situation the Lorentz boost is a rotational one and not a
translational one.

\section{Conclusions}
\label{sec.5}

Cl\'ement~\cite{CL1}, Hirschmann and Welch~\cite{HW} and Cataldo and Salgado~\cite{Cat_Sal}
have found the static magnetic counterpart of the static electric BTZ
black hole~\cite{btz_PRL,btz_PRD}.  This static magnetic solution is
horizonless and has a conical singularity at the origin.  In this
paper we have extended their work in order to include angular
momentum, the definition of conserved quantities and a new
interpretation for the magnetic field source.  We have shown that both
the static and rotating magnetic solutions have negative mass and that
there is an upper bound for the intensity of the magnetic field
strength and for the value of the angular momentum.  Our rotating
magnetic solution is the counterpart of the rotating electric BTZ
black hole~\cite{BTZ_Q} and, as expected, our solution reduces to the
rotating uncharged BTZ solution when the magnetic field source
vanishes.

Hirchmann and Welch~\cite{HW} interpreted the static magnetic source
as a kind of magnetic monopole reminiscent of a Nielson-Oleson vortex
solution.  We prefer to interpret the static magnetic field source as
being composed by a system of two symmetric and superposed electric
charges. One of the electric charges is at rest and the other is
spinning. This system produces no electric field since the total
electric charge is zero and the scalar magnetic field is produced by
the angular electric current.  When we apply a rotational Lorentz
boost to add angular momentum to the spacetime, there appears an
electric charge and an electric field.

\acknowledgments 

This work was partially funded by Funda\c c\~ao para a Ci\^encia e
Tecnologia (FCT) through project CERN/FIS/43797/2001 and
PESO/PRO/2000/4014.  OJCD also acknowledges finantial support from the
portuguese FCT through PRAXIS XXI programme. JPSL thanks
Observat\'orio Nacional do Rio de Janeiro for hospitality.

\end{document}